# Orientation independent quantification of macromolecular proton fraction in tissues with suppression of residual dipolar coupling

Zijian Gao[1], Ziqiang Yu[1], Ziqin Zhou[1,2], Jian Hou[1], Baiyan Jiang[1,3], Michael Ong[4], Weitian Chen[1]

1. Department of Imaging and Interventional Radiology, The Chinese University of Hong Kong, Hong Kong.

2. MR Collaboration, Siemens Heathineers Ltd., Hong Kong.

3. Illumination Medical Technology Limited, Hong Kong.

4. Department of Orthopaedics and Traumatology, The Chinese University of Hong Kong, Hong Kong.

*Corresponding author:

Weitian Chen

Department of Imaging & Interventional Radiology

The Chinese University of Hong Kong

Shatin, Hong Kong, SAR

(852)-3505-1036

Email: wtchen@cuhk.edu.hk

**This PDF file includes:**

    Main Text
    Figures 1 to 7
    Tables 1

**Abstract**

Quantitative magnetization transfer (MT) imaging enables non-invasive characterization of the macromolecular environment of tissues. However, recent work has highlighted that the quantification of MT parameters using saturation radiofrequency (RF) pulses exhibits orientation dependence in ordered tissue structures, potentially confounding its clinical applications. Notably, in tissues with ordered structures, such as articular cartilage and myelin, the residual dipolar coupling (RDC) effect can arise owing to incomplete averaging of dipolar–dipolar interactions of water protons. In this study, we demonstrated the confounding effect of RDC on quantitative MT imaging in ordered tissues can be suppressed by using an emerging technique known as macromolecular proton fraction mapping based on spin-lock (MPF-SL). The off-resonance spin-lock RF pulse in MPF-SL could be designed to generate a strong effective spin-lock field to suppress RDC without violating the specific absorption rate and hardware limitations in clinical scans. Furthermore, suppressing the water pool contribution in MPF-SL enabled the application of a strong effective spin-lock field without confounding effects from direct water saturation. Our findings were experimentally validated using human knee specimens and healthy human cartilage. The results demonstrated that MPF-SL exhibits lower sensitivity to tissue orientation compared with $R_2$, $R_{1\rho}$, and saturation-pulse based MT imaging. Consequently, MPF-SL could serve as a valuable orientation-independent technique for the quantification of MPF.

# 1. INTRODUCTION

The orientational anisotropy of tissues with ordered structures often confounds the quantification of tissue parameters in magnetic resonance imaging (MRI). Owing to this anisotropy, different MRI signal intensities are observed at different orientations of ordered tissue structures relative to the static magnetic field $B_0$. Articular cartilage and myelin, representative ordered tissue structures in the human body, consist of a network of macromolecular fibers [1,2]. The motion of water molecules within these tissues is restricted by the spatial arrangement of these fibers. Each hydrogen nucleus generates a local dipolar field, resulting in the dipolar–dipolar interaction vector <H–H> between neighbouring nuclei [3–5]. In ordered tissue structures, the orientation of restricted water molecules aligns with that of macromolecular fibers, leading to incomplete spatial averaging of the dipolar–dipolar interaction. This phenomenon is known as residual dipolar coupling (RDC) and the residual dipolar-dipolar interaction vector <H-H> approximately aligns with the orientation of macromolecular fibers (Figure 1). Specifically, the influence of RDC varies with the orientation angle $\theta$ between the ordered tissue structure and static magnetic field $B_0$, reaching its minimum value at a "magic angle" ($\theta \approx 54.7°$). Overall, orientational anisotropy results in the orientation dependence of MRI signals owing to RDC, leading to the well-known "magic-angle effect" [6,7], commonly observed in $T_2$ maps [8–10].

The application of strong spin-lock radiofrequency (RF) pulses has been shown to effectively suppress RDC and yield orientation-independent MRI signals[11,12]. Leveraging spin-lock RF pulses, the spin-lattice relaxation time at the rotating frame, also known as $T_{1\rho}$, can be measured[13]. $T_{1\rho}$ provides valuable insights into low-frequency motion and biochemical properties in various biological tissues, including the musculoskeletal system, intervertebral discs, and the brain [14–17]. When the amplitude of the spin-lock RF pulse is significantly larger than the local dipolar field of the nuclei, the secular part of the spectral density of the spin-lattice relaxation rate in a rotating frame ($R_{1\rho} = 1/T_{1\rho}$) can be minimized, thereby suppressing the contribution from dipolar-dipolar interaction to relaxation. Notably, Akella et al.[11] demonstrated the suppression of RDC in cartilage using strong spin-lock RF pulses. Additionally, Casula[18] and Hanninen et al.[19] investigated adiabatic $T_{1\rho}$ pulses, noted to be independent of tissue orientation. Pang[20] proposed a novel order parameter for cartilage measurements based on dispersion fitting. However, despite its potential in various applications, $T_{1\rho}$ lacks specificity for tissue characterization and can be influenced by multiple tissue parameters in vivo[21].

Quantification of magnetization transfer (MT) parameters, such as the macromolecular proton fraction (MPF), can facilitate the measurement of the macromolecular environment of tissues, offering valuable insights into their biochemical composition and molecular properties[22–24]. Prior studies have highlighted the orientation dependence of MT parameters in white matter, with quantitative MT commonly performed based on saturation RF pulses[25,26]. Notably, the widely used two-pool model for MT does not account for the RDC condition observed in ordered tissues. In particular, the line shape (e.g., Gaussian or super-Lorentzian) of the two-pool model does not exhibit any anisotropy[27]. Although Pampel et al. proposed a postprocessing method to correct the two-pool model for the RDC condition, this method requires the acquisition of diffusion-weighted images[26].

Recently, a novel technique, named macromolecular proton fraction mapping based on spin-lock (MPF-SL), has been proposed for quantitative MT based on off-resonance spin-lock RF pulse[28]. However, the properties of MPF-SL in applications involving ordered tissue structures require further investigation. The use of spin-lock RF pulses can facilitate RDC suppression when the spin-lock field is sufficiently strong, thereby alleviating the orientation dependence of MRI signals. However, for on-resonance spin-lock RF pulse, the amplitude of the spin-lock field is typically less than 1000 Hz owing to limitations of the specific absorption rate (SAR) and the power of the RF amplifier. Thus, orientation dependence is often observed in on-resonance $T_{1\rho}$ imaging with spin-lock field < 1000 Hz. In contrast, in the case of off-resonance spin-lock pulse, the spin-lock field is a combination of the $B_1$ field from RF pulses and resonance frequency offset, enabling the realization of a strong spin-lock field without violating the SAR and hardware limitations.

Furthermore, the MPF-SL technique suppresses the contribution of the water pool by utilizing the difference of two $R_{1\rho}$ measured at the same spin-lock angle but with different spin-lock field strength, enabling the use of large RF amplitudes without quantification errors from direct water saturation. Consequently, off-resonance spin-lock based MPF-SL can enable the realization of orientation-independent quantitative MT imaging.

Considering these aspects, in this study, we explored the orientation independence of MPF mapping using MPF-SL. The observations were validated through experiments involving human knee specimens and in vivo human MRI scans.

## 2. THEORY AND METHODS

### 2.1 Quantitative magnetization transfer

Magnetization transfer is an NMR phenomenon in which spins in two or more environments exchange their magnetization. The two-pool model is widely used to describe this phenomenon with a free water pool (A pool) and a bound water pool (B pool). There are six unknown parameters in the two-pool model: MPF, exchange rate, longitudinal relaxation time of A pool and B pool ($T_{1a}$ and $T_{1b}$), and transverse relaxation time of A pool and B pool ($T_{2a}$ and $T_{2b}$). The exchange rate is subdivided into $k_{ba}$ and $k_{ab}$, noticing the exchange rate from B pool to A pool and the exchange rate from A pool to B pool, respectively.

The off-resonance saturation-pulse based MT approach with spoiled gradient sequence is commonly applied to quantify the MT parameters. Z-spectroscopic experiment was originally used to fit multiple MT parameters with MT-weighted images and an independent $T_1$ map. However, Z-spectroscopic experiment requires time-consuming data acquisition and complicated post-processing[29–31]. Single-point MPF mapping method has been developed to address this issue and is used in this study[32]. The source data of the single-point MPF mapping method includes a MT-weighted image, a reference image without MT saturation, an independent $T_1$ map, a $B_0$ map and a $B_1$ map.

**2.2 MPF-SL Method**

MPF-SL is a novel method for MPF quantification using off-resonance spin-lock RF pulses. Considering the two-pool model, the $R_{1\rho}$ relaxation rate can be derived by solving the Bloch-McConnell equation:

$$R_{1\rho} = R_w(\Delta\omega, \omega_1) + R_{mt}(\Delta\omega, \omega_1) \tag{1}$$

where $R_w$ and $R_{mt}$ are associated to the effective relaxation rates in the rotating frame from the free water pool and MT pool, respectively; $\Delta\omega$ is the resonance frequency offset; and $\omega_1$ is the $B_1$ amplitude of the spin-lock RF pulse or frequency of spin-lock (FSL).

$R_w$ is solved by the analytical solution of the Bloch-McConnell equation:

$$R_w(\Delta\omega, \omega_1) = R_{1a} \cos^2\varphi + R_{2a} \sin^2\varphi \tag{2}$$

where $\cos^2\varphi = \frac{\omega_1^2}{\omega_1^2 + \Delta\omega^2}$, $\sin^2\varphi = \frac{\Delta\omega^2}{\omega_1^2 + \Delta\omega^2}$, $R_{1a} = 1/T_{1a}$, and $R_{2a} = 1/T_{2a}$.

If we acquire $R_{1\rho}^{(1)}$ with acquisition parameter $\Delta\omega^{(1)}$ and $\omega_1^{(1)}$, and $R_{1\rho}^{(2)}$ with acquisition parameter $\Delta\omega^{(2)}$ and $\omega_1^{(2)}$ under the condition $\Delta\omega^{(1)}/\omega_1^{(1)} = \Delta\omega^{(2)}/\omega_1^{(2)}$, we can obtain a relaxation rate variable which is specific to MT pool[28]:

$$\begin{aligned} R_{mpfsl} &= R_{1\rho}^{(2)} - R_{1\rho}^{(1)} = \Delta R_{mt} \\ &= k_{ba}^2 f_b (1+f_b) \left[ \left( \frac{1}{(1+f_b)k_{ba} + R_{rfc}^{(1)}} \right) - \left( \frac{1}{(1+f_b)k_{ba} + R_{rfc}^{(2)}} \right) \right] \end{aligned} \tag{3}$$

where the saturation rate $R_{rfc} = \omega_1^2 \pi g(\Delta\omega)$ with Super-Lorentzian line-shape for living tissues[33], and $f_b$ is the pool population ratio of MT pool and the MPF regards as $f = f_b/(1+f_b)$. Note that Eq. (3) is derived under the condition $\Delta\omega^{(1)}/\omega_1^{(1)} \gg 1$ and $\Delta\omega^{(1)} \gg R_{2a}$, which are valid under common acquisition parameters. Moreover, the influence of chemical exchange in Eq. (3) can be neglected at the resonance frequency offsets typically used in MPF-SL [28]

Instead of acquiring enough data to measure $R_{1\rho}^{(1)}$ and $R_{1\rho}^{(2)}$, in reference[28], a method using 180° inversion toggling RF pulse combined with an off-resonance spin-lock RF pulses were used to obtain a value approximate to $R_{mpfsl}$, which can be converted to MPF with knowledge of a $B_1$ map[34].

**2.3 MPF-SL of ordered tissue structures with residual dipolar coupling**

Notably, in ordered tissue structures, $R_{1\rho}$ has anisotropy contribution due to the RDC of water hydrogen protons. However, this is not considered in Bloch-McConnell equations used to derive Eq. (1-2). In ordered tissue, the relaxation rate $R_{1\rho}$ under off-resonance spin-lock field can be expressed as

$$R_{1\rho} = R_w^i + R_w^{ani}(\theta) + R_{mt}(\Delta\omega, \omega_1) \quad (4)$$

where $R_w^i$ and $R_w^{ani}(\theta)$ denote the relaxation rates of the water pool, corresponding to isotropic and anisotropic water molecular relaxation, respectively; $R_{mt}$ is the relaxation rate owing to the MT effect; and $\theta$ is the orientation of ordered tissue structure with respect to the static magnetic field $B_0$.

The anisotropic water molecular relaxation rate $R_w^{ani}(\theta)$ is expressed as[20]

$$R_w^{ani}(\theta) = \frac{R_2^{ani}(\theta)}{1+4\omega_{eff}^2\tau_b^2} \quad (5)$$

where $\omega_{eff}$ is the strength of the effective spin-lock field at off-resonance [14], which equals $\sqrt{\Delta\omega^2 + \omega_1^2}$; $R_2^{ani}(\theta) = R_2^{ani,m}(3\cos^2\theta - 1)^2/4$; and $R_2^{ani,m}$ is the maximum value of $R_2^{ani}(\theta)$ [20].

Combining the definition of $R_{mpfsl}$ with Eq. (4) and (5) yields:

$$R_{mpfsl} = R_{1\rho}^{(2)} - R_{1\rho}^{(1)} = \Delta(\frac{R_2^{ani}(\theta)}{1+4\omega_{eff}^2\tau_b^2}) + \Delta R_{mt} \qquad (6)$$

The element $\Delta(\frac{R_2^{ani}(\theta)}{1+4\omega_{eff}^2\tau_b^2})$ can be minimized using a strong $\omega_{eff}$, effectively suppressing the RDC effect and resulting in low sensitivity of $R_{mpfsl}$ and MPF quantification to tissue orientation.

**2.4 Experiments setup and analysis**

We conducted experiments on human knee specimens and validated the findings on the knee of a healthy human volunteer. We compared the $R_2$, $R_{1\rho}$, $R_1$, and MPF maps measured using the saturation-pulse based approach [32], as well as the MPF maps measured using MPF-SL. The, $R_2$, $R_{1\rho}$, and $R_1$ maps were obtained using established knee imaging protocols. In the saturation-pulse based approach, the MPF was obtained by fitting a two-pool model using the single-point MPF mapping method [32]. In MPF-SL, a relaxation rate specific to the MT effect, $R_{mpfsl}$, was derived. The MPF was then calculated from $R_{mpfsl}$ using a dictionary constructed using the two-pool model based on Bloch–McConnell equation, incorporating constraints associated with $T_{2b}$ and $k_{ba}$ [34].

**2.4.1 Preparation of specimens and in vivo experiments**

Seven human knee specimens were collected from different people undergoing total knee replacement surgeries and preserved in a 10% formalin solution to maintain their tissue properties. All experiments were conducted under approval from the Institutional Review Board. To ensure stability and proper positioning during imaging, the specimen was affixed to a sealed plastic square container using ethyl-2-cyanoacrylate adhesive (Henkel Ltd, Germany). The container with the specimen was filled with phosphate-buffered saline at room temperature (around 20°C) with pH = 7.2–7.4. The container was attached to a custom device, enabling precise orientational control in the scanner (Figure 2).

A healthy male volunteer, aged 29, was enrolled in this experiment. The MRI scan was conducted on the right knee under approval of the Institutional Review Board.

**2.4.2 Data acquisition**

All investigations were conducted using a 3T MRI Scanner (Prisma, Siemens Healthcare, Germany) equipped with a Tx/Rx Knee Coil. The MRI imaging protocols involved the following parameters: 1) proton density-weighted image with TE = 9.6 ms and TR = 2000 ms. 2) $R_1$ and $R_2$ maps obtained using MapIt (Siemens Healthcare, Germany). 3) $R_{1\rho}$ maps based on on-resonance $T_{1\rho}$-prepared two-dimensional turbo spin echo acquisition with time-of-spin lock (TSL) = 0, 10, 30, and 50 ms and FSL = 500 Hz [35]. 4) MPF-SL acquisition with $\Delta\omega^{(1)} = 2\pi \cdot 1000 \text{ rad}$, $\omega_1^{(1)} = 2\pi \cdot 100 \text{ rad}$, $N = \Delta\omega^{(2)}/\Delta\omega^{(1)} = \omega_1^{(2)}/\omega_1^{(1)} = 5$, and TSL = 60 ms. Four off-resonance $T_{1\rho}$-weighted images were acquired in a single MPF-SL scan. The scan time is comparable between MPF-SL and on-resonance $R_{1\rho}$ imaging. Note the off-resonance spin-lock used in MPF-SL generates sufficiently strong effective spin-lock field (~5000Hz) within SAR safety limits and hardware restriction by utilizing a large resonance frequency offset. In contrast, achieving the same effective spin-lock field for on-resonance $R_{1\rho}$ maps is impossible in clinical MRI systems. 5) Single-point MPF mapping method using an MT-weighted spoiled gradient echo (GRE) with the Gaussian pulse for off-resonance saturation ($B_0$ = 4000 Hz) and effective saturation flip angles (FA = 600°) at 20 ms duration, and no MT-weighted images ($\Delta$ =100 kHz). To validate the postprocessing, in addition to single-point acquisition, we also acquired Z-spectroscopic data with 11 $\Delta$ values in the range of 2–36 kHz with the Gaussian pulse of saturation flip angles of 950° and 600° during the first human knee specimen experiment at an orientation of 0°. The Z-spectroscopic data were used to calculate four parameters: $f$, $k_{ab}$, $T_{2a}$, and $T_{2b}$. 6) $B_1$ and $B_0$ maps collected using the Siemens clinical knee imaging protocol.

All specimen's images were acquired at orientations of 0°, 15°, 30°, 45°, 60°, 75°, and 90° with respect to $B_0$. Each specimen was positioned at the center of the knee transmit/receive (T/R) coil and manually rotated using a custom-built rotation device, allowing precise orientation control. Other imaging parameters for the specimens were set as follows: field of view (FOV) = 110 mm × 110 mm, slice thickness = 3 mm, number of slices 1, and image resolution = 0.4 mm × 0.4 mm. The $B_1$ and $B_0$ maps were acquired at lower resolutions of 1.2 mm × 1.2 mm and 1.8 mm × 1.8 mm, respectively. The scan time to acquire all images at each orientation was approximately 18 minutes, resulting in a total scan time of about 2 hours per specimen. The Z-spectroscopic data took additional approximately 12 minutes to acquire. For the in vivo human knee scan, the FOV was 150 mm × 150 mm and the image resolution was 0.7 mm × 0.7 mm. The resolutions of $B_1$ and $B_0$ maps for the in vivo human scan were 1.4 mm × 1.4 mm and 1.8 mm × 1.8 mm, respectively. The total scan time for the in vivo knee scan was approximately 16 minutes.

**2.4.3 Data analysis**

The single-point MPF map was obtained using reconstruction algorithms with constraints $k_{ab}(1-f)/f$, $T_{2b}$, and $R_{1a}T_{2a}$ [32]. From the Z-spectroscopic data of the two-pool model and T1 map, we obtained $k_{ab}(1-f)/f$ = 35 s$^{-1}$, $T_{2b}$ = 6 μs, and $R_{1a}T_{2a}$ = 0.035.

$R_{mpfsl}$ maps were calculated from four magnetization-prepared images. Detailed information regarding the calculation of $R_{mpfsl}$ can be found in [28]. The dictionary was generated using $T_{2b}$ = 6 μs and $k_{ba}$ = 45 s$^{-1}$ measured from Z-spectroscopic data. The $B_1$ range for generating the dictionary was 0.8–1.3.

In the specimen experiments, two regions of interest (ROIs) were manually drawn on the cartilage at all seven orientations. ROI1 included mostly the superficial and middle zones of the cartilage and was close to the surface, while ROI2 mostly constituted the deep and middle zones of the cartilage. In the in vivo human scan, angle-based ROIs were manually drawn by assuming a circular cartilage shape at 4-degree orientation intervals[36].

The anisotropy rate [37] quantifies the orientation dependence of a measurement as

$$Anisotropy\ rate = \frac{R^{max} - R^{min}}{R^{max} + R^{min}} \quad (7)$$

where $R^{max}$ and $R^{min}$ denote the maximum and minimum relaxation rates measured across all orientations, respectively. The anisotropy rate can serve as a performance standard, with low and high values indicating orientation independence and orientation dependence, respectively.

All data analyses were performed using custom-written code in MATLAB R2023b (MathWorks, USA), except for Z-spectroscopic analysis, which was performed using the qMRLab opensource tool [38].

## 3. RESULTS

### 3.1 Specimen experiments

Figure 3 shows the results from the first human knee specimen (S1). In the MPF-SL technique, a relaxation rate, $R_{mpfsl}$, related to the MT effect is calculated, from which the MPF is derived. Note $R_2 = 1/T_2$ and $R_{1\rho}$ exhibit variations with the tissue orientation, consistent with prior work [7]. In contrast, $R_1$ and $R_{mpfsl}$ are near independent of orientation. The MPF obtained using MPF-SL ($MPF_{SL}$) demonstrates less sensitivity to orientation compared with the MPF measured using the saturation-pulse based approach ($MPF_{ST}$). Two ROIs are selected for analysis, as shown in Figure 4(a). The relaxation rates in these two ROIs are calculated at different orientation angles (0°, 15°, 30°, 45°, 60°, 75°, and 90° with respect to $B_0$) (Figures 4(c–h)). The anisotropy rate, as described in the Data Analysis section, is calculated to quantify the orientation dependence of the measured parameters (Figure 4(b)). $R_{mpfsl}$ exhibits significantly lower sensitivity to orientation compared with $R_2$ and $R_{1\rho}$. Moreover, compared with $MPF_{ST}$, $MPF_{SL}$ shows significantly reduced orientation dependence in MPF map. Similar results are observed for the other human knee specimens (S2-S7), as outlined in the Supplementary Information (Figures S1–S12). Table 1 summarizes the anisotropy rate of all specimens.

Figure 5 presents the statistical analysis of anisotropy rate of the measured parameters from all specimens. The group differences analysis was applied to anisotropy rate, utilizing Analysis of Variance (ANOVA) testing to calculate P value and mean difference (D). Results indicate that there is a significant difference between the anisotropy rate of $R_{mpfsl}$ and that of $R_2$ (P < 0.05, D=38.82), as well as between the anisotropy rate of $R_{mpfsl}$ and that of $R_{1\rho}$ (P < 0.05, D=18.64). On the contrary, the difference between the anisotropy rate of $R_{mpfsl}$ and that of $R_1$ is not significant (P = 0.786, D=3.78). Moreover, a significant difference exists between the anisotropy rate of $MPF_{SL}$ and that of $MPF_{ST}$ (P < 0.05, D=11.34).

## 3.2 In vivo experiments

To further investigate the RDC effect on quantitative MRI in vivo, a knee MRI experiment was conducted on a healthy human volunteer. Maps of $R_2$, $R_{1\rho}$, $R_1$, $R_{mpfsl}$, $MPF_{ST}$, and $MPF_{SL}$ of the cartilage are segmented and displayed alongside anatomic images in Figure 6. The angle-based ROIs are drawn on the cartilage, with the angle between the cartilage surface and $B_0$ ranging from -90° to +90°[36], as shown in Figure 7(a). The means ± standard deviation of the measured parameters within ROIs reflect the orientation dependence of these parameters (Figures 7(c–h)). Figure 7(b) shows the anisotropy rate of these parameters. Notably, $R_2$ and $R_{1\rho}$ exhibit higher anisotropy rates (~41% and ~27%, respectively) compared with $R_1$ and $R_{mpfsl}$ (~14% and ~15%, respectively). $MPF_{SL}$ exhibits significantly reduced orientation dependence, with an anisotropy rate of approximately 14%, compared with $MPF_{ST}$ (anisotropy rate of approximately 22%).

## 4. DISCUSSION

Our specimen and in vivo studies highlight that off-resonance spin-lock based MPF-SL can effectively suppress the RDC effect in the quantification of MPF in ordered tissue structures. In contrast, quantification using the conventional saturation-pulse based approach may be confounded by RDC. The following discussion explores the mechanisms underlying these observations and analyzes our experimental setup.

**4.1 Residual dipolar coupling in MPF-SL and saturation-pulse based approach**

In nuclear magnetic resonance theory, each proton generates a local dipolar magnetic field that interacts with the local field of its neighbouring protons [3,5,39,40]. In ordered tissues, the structured microarchitecture prevents complete averaging of dipolar-dipolar interactions, resulting in RDC. This RDC contributes to anisotropic relaxation rates. The Hamiltonian describing this system can be approximated as follows [27,41]:

$$H = H_z + \overline{H_D} + H(t) \tag{8}$$

$$H(t) = H_D(t) - \overline{H_D} \tag{9}$$

where $H_z$ denotes Zeeman interaction, $H_D(t)$ is the time-dependent dipolar-dipolar interaction, and $\overline{H_D}$ is the time average of dipolar-dipolar interaction. RDC arises when $\overline{H_D}$ is non-zero. Notably, $\overline{H_D}$ is associated with the angle $\theta_d$ between the external magnetic field and the direction of dipolar–dipolar interaction:

$$\overline{H_{D,\theta}} = \overline{H_{D,\theta=0}} (\frac{3\cos^2\theta_d - 1}{2}) \tag{10}$$

The dipolar-dipolar interaction $\overline{H_{D,\theta}}$ between protons diminishes when the angle between the external magnetic field and the internuclear vector (dipolar-dipolar interaction vector) is approximately 54.7°, known as the "magic angle." In ordered tissue, such as cartilage, the movement of water protons is restricted by the matrix of macromolecular structures, allowing the RDC to be measured in the MR signal. When the amplitude of the rotating RF field is considerably greater than the local dipolar magnetic field, the secular part of the spectral density of the spin-lattice relaxation rate in a rotating frame can be minimized, thereby suppressing the contribution from dipolar-dipolar interaction to the relaxation [11–13]. Thus, a spin-lock pulse with high amplitude (>> 1000 Hz) can be used to achieve orientation-independent imaging [11]. In the case of MPF-SL, off-resonance spin-lock pulse can be used to achieve a strong spin-lock field without violating the SAR and RF power limits, resulting in lower sensitivity to tissue orientation.

MT parameters are typically quantified using saturation-pulse based approach. The single-point MPF mapping method used in this study is a state-of-the-art saturation-pulse based method and is considered the fastest approach for MPF mapping [32]. However, saturation-pulse based approach may not fully account for the presence of anisotropic interactions in ordered tissues. The widely accepted super-Lorentzian line-shape in the MT model [27] does not explicitly incorporate anisotropy effects arising from dipolar–dipolar interaction.

Notably, the magnetizations under the saturation-pulse based and spin-lock based approach the same steady-stage magnetization. The RDC effect can potentially be mitigated in saturation-pulse based approach by increasing the amplitude of the saturation RF pulses. However, this leads to intensified direct water saturation effects in saturation-pulse based approach. In contrast, increasing the amplitude of the spin-lock RF pulse used in MPF-SL does not induce the adversarial direct water saturation effect because the signal from the water pool is eliminated in MPF-SL.

**4.2 Choice of parameters for MPF-SL**

Achieving optimal parameters for MPF-SL to ensure a robust $R_{mpfsl}$ signal level while maintaining measurement independence from tissue orientation requires a balance between signal strength and the limitations of SAR and RF hardware. In MPF-SL, we collect data with two sets of RF parameters ($\Delta\omega_1^{(1)}$, $\omega_1^{(1)}$) and ($\Delta\omega_1^{(2)}$, $\omega_1^{(2)}$), and a scaling factor $N$ under the condition $\Delta\omega^{(1)}/\omega_1^{(1)} = \Delta\omega^{(2)}/\omega_1^{(2)}$. Considering SAR and RF power limitations, the FSL typically remains below 500 Hz. In MPF-SL, the resonance frequency offset is typically chosen such that $\Delta\omega \gg \omega_1$, thereby avoiding signal contamination from chemical exchange and the nuclear Overhauser effect. In the context of RDC suppression, a small resonance frequency offset leads to a greater contribution of $R_2$ relaxation to the signal, necessitating a higher $B_1$ RF field to suppress RDC under the limitations of SAR and RF hardware. A large resonance frequency offset results in higher effective spin-lock field and superior suppression of RDC. However, further increasing the resonance frequency offset causes $R_{mpfsl}$ to approach $R_1$ and a diminished MT signal in $R_{mpfsl}$. Increased resonance frequency offsets also increase the sensitivity of $R_{mpfsl}$ to variations in the $T_2$ relaxation of the MT pool, which is undesirable for MPF-SL as the $T_2$ of the MT pool is assumed to be constant in MPF-SL. Under these considerations, we set $\Delta\omega^{(1)} = 2\pi \cdot 1000$ rad, $\omega_1^{(1)} = 2\pi \cdot 100$ rad, and $N = 5$ in this study.

**4.3 limitation and challenges**

Despite the promising results, our study has several limitations and challenges that warrant further investigation. 1) In the specimen experiments, orientation control relies on manual rotation using a hand-made device. Utilizing a more precise device could potentially improve orientation control. Additionally, the specimens were obtained from total knee replacement surgeries in elderly patients. The microstructural changes in cartilage relative to human age were not considered in this study, which is a potential confounding factor[42]. 2) For in vivo experiments, using advanced orientation imaging techniques such as diffusion tensor imaging[43] could provide more accurate orientation references compared to the manually selected ROI method. By using orientation imaging, the orientation-dependency of quantification of magnetization transfer parameters in vivo can be validated more reliably. 3) While our method shows potential for cartilage assessment, further validation is required. Specifically, the relationship between $R_{mpfsl}$ and cartilage composition (e.g., collagen, proteoglycan, and water content associated with chondrocytes) needs thorough investigation. This validation can be conducted by comparing our MRI technique with histological analysis on specimens. 4) Our study lacks experiments demonstrating the clinical utility of this method. Further clinical studies are needed to elucidate the correlation between $R_{mpfsl}$ and various aspects of cartilage health, including degenerative changes, traumatic injuries, and osteoarthritis.

## 5. CONCLUSION

Common saturation-pulse based approach for measuring magnetization transfer parameters can be affected by residual dipolar coupling in ordered tissue structures, leading to orientation-dependent results that may complicate clinical diagnoses. In this study, we demonstrate that this confounding effect can be suppressed using the recently proposed spin-lock based quantitative magnetization transfer imaging technique, applied to human knee specimens experiments and validated in vivo on healthy human experiments. This innovative technique shows promising potential for the accurate characterization of ordered tissues, particularly in structures such as cartilage and myelin, potentially enhancing the reliability of MT-based diagnostic imaging.

## ABBREVIATIONS USED

| | |
|---|---|
| FA | flip angles |
| FOV | field of view |
| MPF | macromolecular proton fraction |
| MT | magnetization transfer |
| RDC | residual dipolar coupling |
| RF | radiofrequency |
| ROI | region of interest |
| SAR | specific absorption rate |
| SL | spin-lock |
| ST | saturation |
| TSL | time of spin lock |

## ACKNOWLEDGEMENT

This study was supported by a grant from the Research Grants Council of the Hong Kong SAR (Project GRF 14213322), and a grant from the Innovation and Technology Commission of the Hong Kong SAR (Project No. MRP/046/20x). We would like to thank Ms. Qianxue Shan for providing valuable feedback.

## DATA AVAILABILITY STATEMENT

Our data that supports the findings of this study are available in the supplementary material of this article.

## REFERENCES

1. Jeffery AK, Blunn GW, Archer CW, Bentley G. Three-dimensional collagen architecture in bovine articular cartilage. *J Bone Joint Surg Br*. 1991;73-B(5):795-801. doi:10.1302/0301-620x.73b5.1894669

2. Rieppo J, Hallikainen J, Jurvelin JS, Kiviranta I, Helminen HJ, Hyttinen MM. Practical considerations in the use of polarized light microscopy in the analysis of the collagen network in articular cartilage. *Microsc Res Tech*. 2008;71(4):279-287. doi:10.1002/jemt.20551

3. Bloembergen N. Spin Relaxation Processes in a Two-Proton System. *Phys Rev*. 1956;104(6):1542-1547. doi:10.1103/PhysRev.104.1542

4. Lipsitz RS, Tjandra N. Residual dipolar couplings in NMR structure analysis. *Annu Rev Biophys Biomol Struct*. 2004;33:387-413. doi:10.1146/annurev.biophys.33.110502.140306

5. Karjalainen J, Henschel H, Nissi MJ, Nieminen MT, Hanni M. Dipolar Relaxation of Water Protons in the Vicinity of a Collagen-like Peptide. *J Phys Chem B*. 2022;126(13):2538-2551. doi:10.1021/acs.jpcb.2c00052

6. Erickson SJ, Prost RW, Timins ME. The "magic angle" effect: background physics and clinical relevance. *Radiology*. 1993;188(1):23-25. doi:10.1148/radiology.188.1.7685531

7. Shao H, Pauli C, Li S, et al. Magic angle effect plays a major role in both T1rho and T2 relaxation in articular cartilage. *Osteoarthritis Cartilage*. 2017;25(12):2022-2030. doi:10.1016/j.joca.2017.01.013

8. Mosher TJ, Smith H, Dardzinski BJ, Schmithorst VJ, Smith MB. MR imaging and T2 mapping of femoral cartilage: in vivo determination of the magic angle effect. *Am J Roentgenol*. 2001;177(3):665-669.

9. Wang L, Regatte RR. Investigation of regional influence of magic-angle effect on T2 in human articular cartilage with osteoarthritis at 3 T. *Acad Radiol*. 2015;22(1):87-92.

10. Leskinen HPP, Hänninen NE, Nissi MJ. $T_2$ orientation anisotropy mapping of articular cartilage using qMRI. *Phys Med Biol*. 2023;68(8):085004. doi:10.1088/1361-6560/acc169

11. Akella SVS, Regatte RR, Wheaton AJ, Borthakur A, Reddy R. Reduction of residual dipolar interaction in cartilage by spin-lock technique. *Magn Reson Med*. 2004;52(5):1103-1109. doi:10.1002/mrm.20241

12. Lee M, Goldburg WI. Nuclear-Magnetic-Resonance Line Narrowing by a Rotating rf Field. *Phys Rev*. 1965;140(4A):A1261-A1271. doi:10.1103/PhysRev.140.A1261

13. Jones GP. Spin-Lattice Relaxation in the Rotating Frame: Weak-Collision Case. *Phys Rev*. 1966;148(1):332-335. doi:10.1103/PhysRev.148.332

14. Wáng YXJ, Zhang Q, Li X, Chen W, Ahuja A, Yuan J. T1ρ magnetic resonance: basic physics principles and applications in knee and intervertebral disc imaging. *Quant Imaging Med Surg*. 2015;5(6):858.

15. De Mello R, Ma Y, Ji Y, Du J, Chang EY. Quantitative MRI Musculoskeletal Techniques: An Update. *Am J Roentgenol*. 2019;213(3):524-533. doi:10.2214/AJR.19.21143

16. Wheaton AJ, Borthakur A, Kneeland JB, Regatte RR, Akella SVS, Reddy R. In vivo quantification of $T_{1\rho}$ using a multislice spin-lock pulse sequence. *Magn Reson Med*. 2004;52(6):1453-1458. doi:10.1002/mrm.20268

17. Menon RG, Sharafi A, Windschuh J, Regatte RR. Bi-exponential 3D-T1ρ mapping of whole brain at 3 T. *Sci Rep*. 2018;8(1):1176. doi:10.1038/s41598-018-19452-5

18. Casula V, Autio J, Nissi MJ, et al. Validation and optimization of adiabatic T1ρ and T2ρ for quantitative imaging of articular cartilage at 3 T. *Magn Reson Med*. 2017;77(3):1265-1275. doi:10.1002/mrm.26183

19. Hänninen N, Rautiainen J, Rieppo L, Saarakkala S, Nissi MJ. Orientation anisotropy of quantitative MRI relaxation parameters in ordered tissue. *Sci Rep*. 2017;7(1):9606. doi:10.1038/s41598-017-10053-2

20. Pang Y. An order parameter without magic angle effect (OPTIMA) derived from dispersion in ordered tissue. *Magn Reson Med*. 2020;83(5):1783-1795. doi:10.1002/mrm.28045

21. Chen W. Errors in quantitative T1rho imaging and the correction methods. *Quant Imaging Med Surg*. 2015;5(4):583-591. doi:10.3978/j.issn.2223-4292.2015.08.05

22. Kisel AA, Naumova AV, Yarnykh VL. Macromolecular proton fraction as a myelin biomarker: principles, validation, and applications. *Front Neurosci*. 2022;16:819912.

23. Sritanyaratana N, Samsonov A, Mossahebi P, Wilson JJ, Block WF, Kijowski R. Cross-relaxation imaging of human patellar cartilage in vivo at 3.0T. *Osteoarthritis Cartilage*. 2014;22(10):1568-1576. doi:10.1016/j.joca.2014.06.004

24. Yarnykh VL, Tartaglione EV, Ioannou GN. Fast macromolecular proton fraction mapping of the human liver in vivo for quantitative assessment of hepatic fibrosis. *NMR Biomed*. 2015;28(12):1716-1725. doi:10.1002/nbm.3437

25. Henkelman RM, Stanisz GJ, Kim JK, Bronskill MJ. Anisotropy of NMR properties of tissues. *Magn Reson Med*. 1994;32(5):592-601. doi:10.1002/mrm.1910320508

26. Pampel A, Müller DK, Anwander A, Marschner H, Möller HE. Orientation dependence of magnetization transfer parameters in human white matter. *NeuroImage*. 2015;114:136-146. doi:10.1016/j.neuroimage.2015.03.068

27. Morrison C, Mark Henkelman R. A Model for Magnetization Transfer in Tissues. *Magn Reson Med*. 1995;33(4):475-482. doi:10.1002/mrm.1910330404

28. Hou J, Wong VW, Jiang B, et al. Macromolecular proton fraction mapping based on spin-lock magnetic resonance imaging. *Magn Reson Med*. 2020;84(6):3157-3171. doi:10.1002/mrm.28362

29. Sled JG, Pike GB. Quantitative interpretation of magnetization transfer in spoiled gradient echo MRI sequences. *J Magn Reson San Diego Calif 1997*. 2000;145(1):24-36. doi:10.1006/jmre.2000.2059

30. Yarnykh VL. Pulsed Z-spectroscopic imaging of cross-relaxation parameters in tissues for human MRI: Theory and clinical applications. *Magn Reson Med*. 2002;47(5):929-939. doi:10.1002/mrm.10120

31. Holt RW, Duerk JL, Hua J, Hurst GC. Estimation of bloch model MT spin system parameters from Z-spectral data. *Magn Reson Med*. 1994;31(2):122-130. doi:10.1002/mrm.1910310205

32. Yarnykh VL. Fast macromolecular proton fraction mapping from a single off-resonance magnetization transfer measurement. *Magn Reson Med*. 2012;68(1):166-178. doi:10.1002/mrm.23224

33. Henkelman RM, Huang X, Xiang QS, Stanisz GJ, Swanson SD, Bronskill MJ. Quantitative interpretation of magnetization transfer. *Magn Reson Med*. 1993;29(6):759-766. doi:10.1002/mrm.1910290607

34. Hou J, Wong VWS, Qian Y, et al. Detecting Early-Stage Liver Fibrosis Using Macromolecular Proton Fraction Mapping Based on Spin-Lock MRI: Preliminary Observations. *J Magn Reson Imaging*. 2023;57(2):485-492. doi:10.1002/jmri.28308

35. Chen W, Chan Q, Wáng YXJ. Breath-hold black blood quantitative T1rho imaging of liver using single shot fast spin echo acquisition. *Quant Imaging Med Surg*. 2016;6(2):168-177. doi:10.21037/qims.2016.04.05

36. Nozaki T, Kaneko Y, Yu HJ, et al. T1rho mapping of entire femoral cartilage using depth- and angle-dependent analysis. *Eur Radiol*. 2016;26:1952-1962. doi:10.1007/s00330-015-3988-5

37. Rieppo J, Hallikainen J, Jurvelin JS, Kiviranta I, Helminen HJ, Hyttinen MM. Practical considerations in the use of polarized light microscopy in the analysis of the collagen network in articular cartilage. *Microsc Res Tech*. 2008;71(4):279-287. doi:10.1002/jemt.20551

38. Karakuzu A, Boudreau M, Duval T, et al. qMRLab: Quantitative MRI analysis, under one umbrella. *J Open Source Softw*. 2020;5(53):2343. doi:10.21105/joss.02343

39. Goldman M. Formal Theory of Spin–Lattice Relaxation. *J Magn Reson*. 2001;149(2):160-187. doi:10.1006/jmre.2000.2239

40. Abragam A. *The Principles of Nuclear Magnetism*. Clarendon Press; 1961.

41. Morrison C, Stanisz G, Henkelman RM. Modeling magnetization transfer for biological-like systems using a semi-solid pool with a super-Lorentzian lineshape and dipolar reservoir. *J Magn Reson B*. 1995;108(2):103-113. doi:10.1006/jmrb.1995.1111

42. Gründer W. MRI assessment of cartilage ultrastructure. *NMR Biomed*. 2006;19(7):855-876. doi:10.1002/nbm.1092

43. Wang N, Mirando AJ, Cofer G, Qi Y, Hilton MJ, Johnson GA. Characterization Complex Collagen Fiber Architecture in Knee Joint Using High Resolution Diffusion Imaging. *Magn Reson Med*. 2020;84(2):908-919. doi:10.1002/mrm.28181

**Figures and Tables**

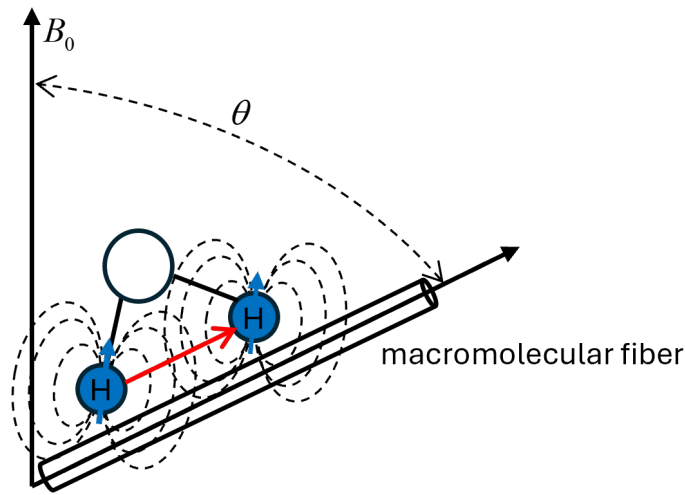

**Figure 1**. The dipolar-dipolar interaction between two water hydrogen protons. $\theta$ is the orientation angle of macromolecular fiber with respect to the static magnetic field $B_0$, the red arrow represents the residual dipolar coupling <H-H> roughly aligned with orientation of macromolecular fiber. The dotted lines represent the local dipolar magnetic field.

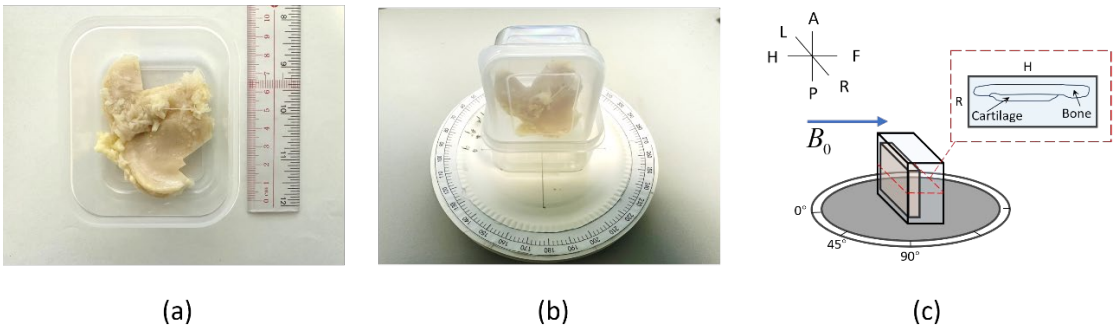

**Figure 2.** (a) Partially completed tibia specimen (S1) fixed in a container. (b) Hand-made rotary device with container and specimen. Orientation of cartilage with respect to $B_0$, controlled by manually adjusting the rotation angle of this device. (c) Schematic of the orientation of specimen during MRI scan. The blue arrow indicates the direction of the static magnetic field. The red dashed box indicates the slice orientation of the acquired image.

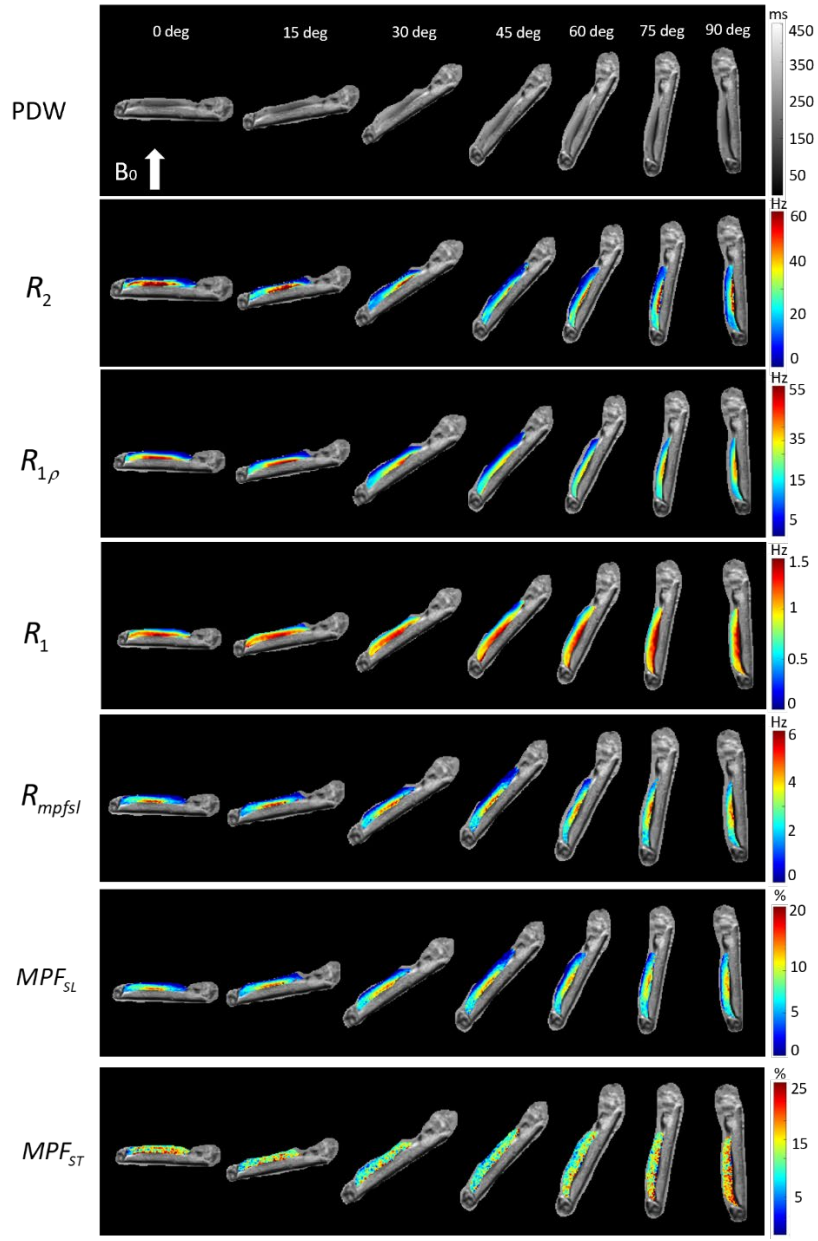

**Figure 3.** Maps of relaxation rates and MPF of knee specimen S1 at different orientations. Top to the bottom: maps of $R_2$, on-resonance $R_{1\rho}$ at FSL 500 Hz, $R_1$, $R_{mpfsl}$, $MPF_{SL}$, and $MPF_{ST}$. Left to right: maps at orientations of 0°, 15°, 30°, 45°, 60°, 75°, and 90° with respect to $B_0$. PDW images are included as anatomical reference images.

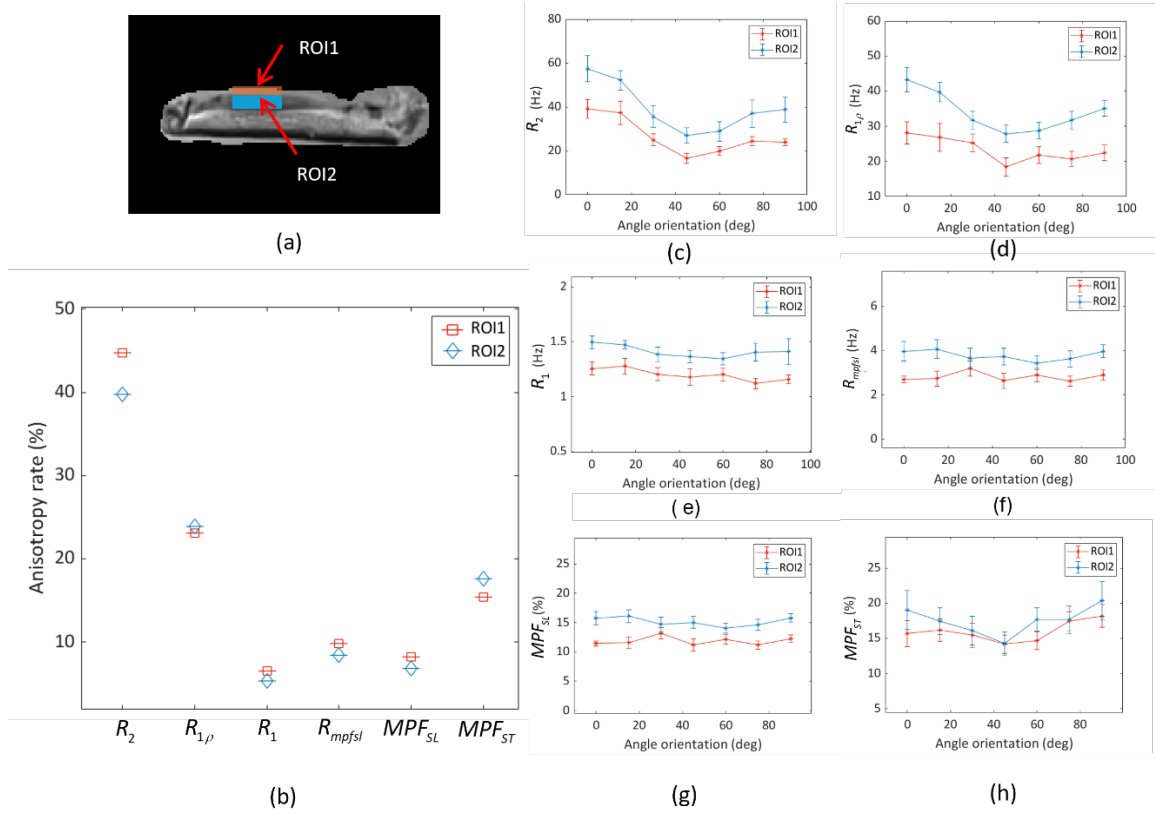

**Figure 4.** (a) Illustration of two ROIs in knee specimen S1. (b) Anisotropy rate of relaxation time maps ( $R_2$, $R_{1\rho}$, $R_1$, and $R_{mpfsl}$ ) and MPF maps ( $MPF_{SL}$ and $MPF_{ST}$ ) in ROI1 and ROI2. (c-h) Mean ± standard

deviation of relaxation rates in ROI1 and ROI2, as shown in (a), were calculated at different angle orientations (c) $R_2$, (d) $R_{1\rho}$, (e) $R_1$, (f) $R_{mpfsl}$, (g) $MPF_{SL}$, and (h) $MPF_{ST}$.

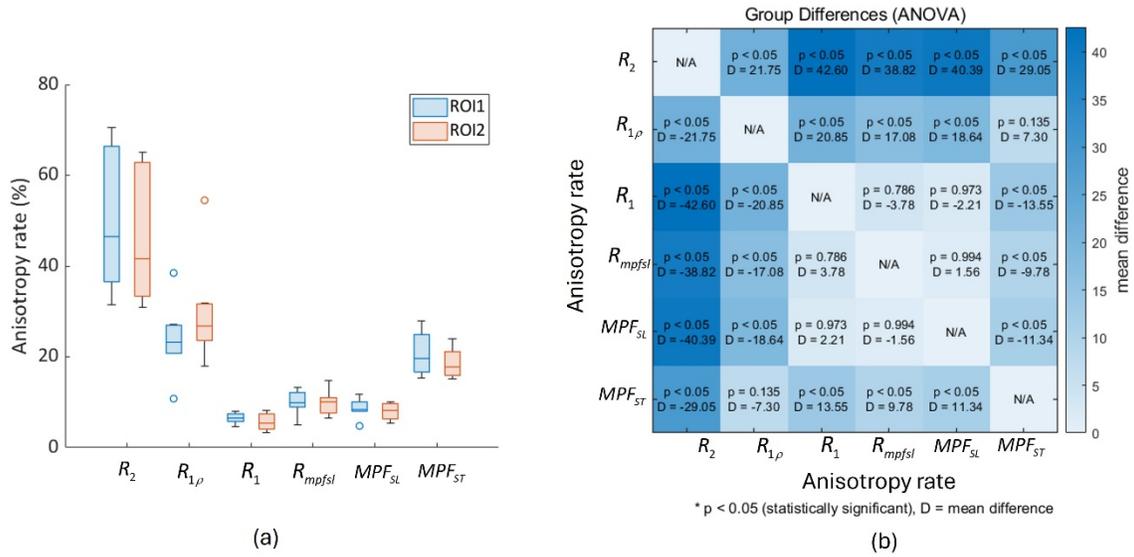

**Figure 5.** Statistics analysis of data from specimen experiments. (a). Box plot comparing anisotropy rate across different methods. (b). Group differences analysis was applied to anisotropy rates of various parameters, utilizing ANOVA to determine P values and mean differences.

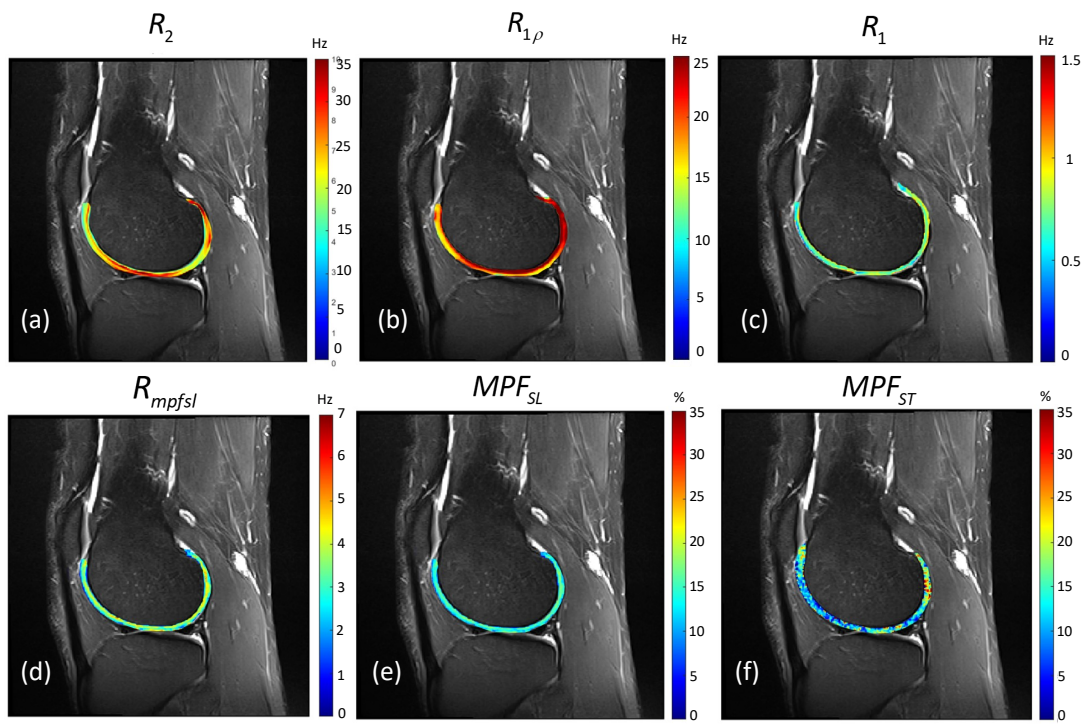

**Figure 6.** Maps of relaxation rates and MPF of cartilage from a healthy volunteer. (a) to (f): maps of $R_2$, on-resonance $R_{1\rho}$ at FSL 500 Hz, $R_1$, $R_{mpfsl}$, $MPF_{SL}$, and $MPF_{ST}$, respectively.

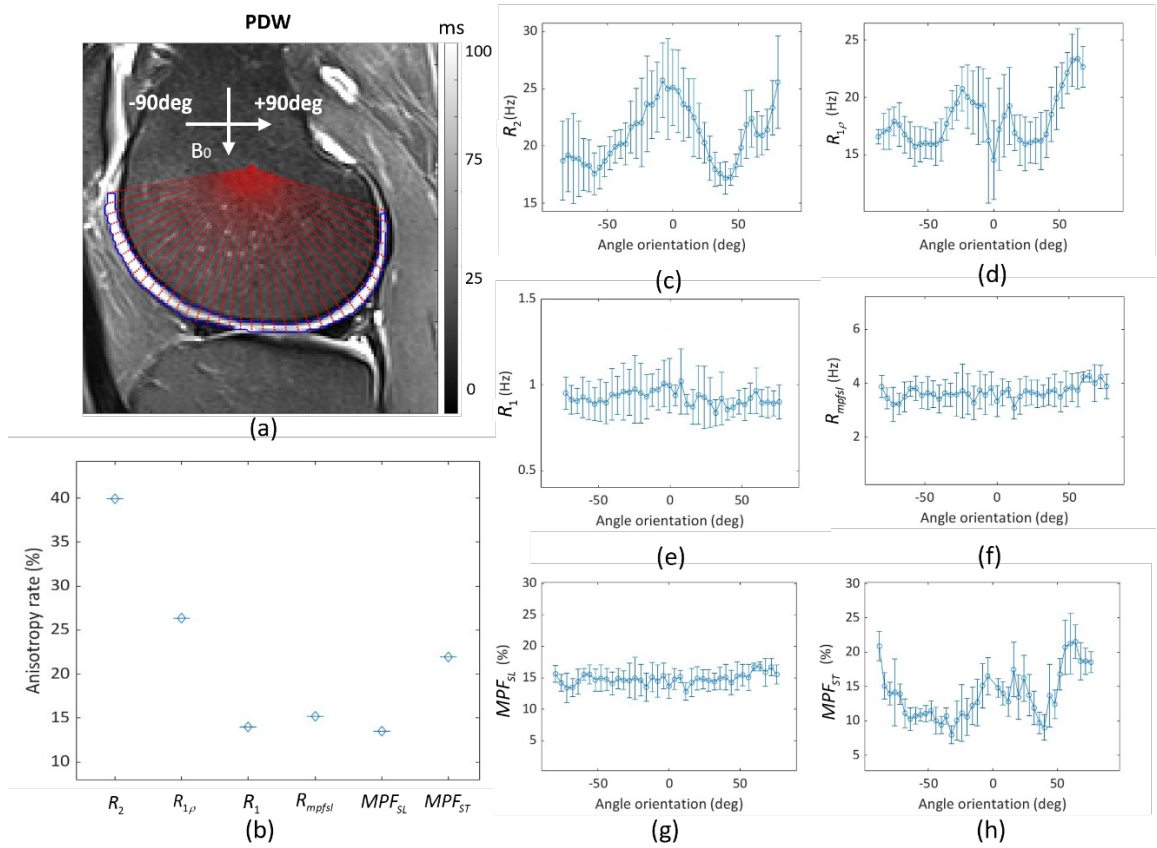

**Figure 7.** (a) Illustration of angle-based ROIs drawn on the cartilage. After manual cartilage segmentation, the angle segmentation in 4-degree increments over the segmented cartilage, with positive angles representing counterclockwise rotation and negative angles representing clockwise rotation. (b) Anisotropy rate of relaxation time maps ($R_2$, $R_{1\rho}$, $R_1$ and $R_{mpfsl}$) and MPF maps ($MPF_{SL}$ and $MPF_{SL}$) calculated from angle-based ROIs. Mean ± standard deviation of relaxation rates and MPF in angle-based ROIs at different angle orientations: (c) $R_2$, (d) $R_{1\rho}$, (e) $R_1$, (f) $R_{mpfsl}$, (g) $MPF_{SL}$, and (h) $MPF_{ST}$.

**Table 1.** Anisotropy rate of relaxation rates and MPF of all specimens (S1–S7).

| ID | Region | $R_2$ | $R_{1\rho}$ | $R_1$ | $R_{mpfsl}$ | $MPF_{SL}$ | $MPF_{ST}$ |
|---|---|---|---|---|---|---|---|
| S1 | ROI1 | 44.69 | 23.17 | 6.72 | 9.81 | 8.22 | 15.39 |
|    | ROI2 | 39.71 | 23.89 | 5.31 | 8.40 | 6.78 | 17.58 |
| S2 | ROI1 | 70.47 | 26.50 | 7.99 | 10.58 | 9.54 | 15.96 |
|    | ROI2 | 65.10 | 31.78 | 6.88 | 9.99 | 8.10 | 17.76 |
| S3 | ROI1 | 70.32 | 38.47 | 7.57 | 12.65 | 11.65 | 18.47 |
|    | ROI2 | 56.77 | 54.47 | 8.12 | 10.86 | 9.86 | 23.86 |
| S4 | ROI1 | 54.21 | 20.98 | 6.51 | 8.64 | 8.43 | 27.85 |
|    | ROI2 | 41.68 | 26.69 | 7.48 | 10.95 | 9.98 | 19.15 |
| S5 | ROI1 | 46.56 | 20.69 | 4.53 | 9.41 | 7.96 | 26.12 |
|    | ROI2 | 64.75 | 31.36 | 5.38 | 6.42 | 5.38 | 21.80 |
| S6 | ROI1 | 33.74 | 10.77 | 6.12 | 4.95 | 4.74 | 19.59 |
|    | ROI2 | 31.30 | 17.89 | 3.26 | 7.34 | 6.02 | 15.38 |
| S7 | ROI1 | 31.41 | 27.21 | 5.65 | 13.29 | 10.30 | 20.91 |
|    | ROI2 | 30.89 | 23.57 | 3.68 | 14.78 | 9.21 | 15.12 |